\documentclass[pre,aps,reprint,superscriptaddress]{revtex4-1}
\usepackage{graphicx}
\usepackage{amsmath,amssymb}
\usepackage{braket}
\usepackage[colorlinks=true,linkcolor=black, citecolor=blue, urlcolor=blue]{hyperref}

\usepackage{soul}

\begin{document}
\title{Randomizing growing networks with a time-respecting null model}

\author{Zhuo-Ming Ren}

\affiliation{College of Economics and Management, Zhejiang University of Technology, Hangzhou 310014, PR  China}
\affiliation{Department of Physics, University of Fribourg, 1700 Fribourg, Switzerland}
\author{Manuel Sebastian Mariani}
\email{manuel.mariani@unifr.ch}
\affiliation{Institute of Fundamental and Frontier Sciences, University of Electronic Science and Technology of China, Chengdu 610054, PR China}
\affiliation{URPP Social Networks, Universit\"at Z\"urich, Switzerland}
\affiliation{Department of Physics, University of Fribourg, 1700 Fribourg, Switzerland}
\author{Yi-Cheng Zhang}
\affiliation{Department of Physics, University of Fribourg, 1700 Fribourg, Switzerland}
\author{Mat{\'u}{\v{s}} Medo}
\email{matus.medo@unifr.ch}
\affiliation{Institute of Fundamental and Frontier Sciences, University of Electronic Science and Technology of China, Chengdu 610054, PR China}
\affiliation{Department of Radiation Oncology, Inselspital, Bern University Hospital and University of Bern, 3010 Bern, Switzerland}
\affiliation{Department of Physics, University of Fribourg, 1700 Fribourg, Switzerland}

\vspace{10pt}

\begin{abstract}
Complex networks are often used to represent systems that are not static but grow with time: people make new friendships, new papers are published and refer to the existing ones, and so forth. To assess the statistical significance of measurements made on such networks, we propose a randomization methodology---a time-respecting null model---that preserves both the network's degree sequence and the time evolution of individual nodes' degree values. By preserving the temporal linking patterns of the analyzed system, the proposed model 
is able to factor out the effect of the system's temporal patterns on its structure. We apply the model to the citation network of Physical Review scholarly papers and the citation network of US movies. The model reveals that the two datasets are strikingly different with respect to their degree-degree correlations, and we discuss the important implications of this finding on the information provided by paradigmatic node centrality metrics such as indegree and Google's PageRank. The randomization methodology proposed here can be used to assess the significance of any structural property in growing networks, which could bring new insights into the problems where null models play a critical role, such as the detection of communities and network motifs. 
\end{abstract}

\maketitle

\section{Introduction}
Complex networks \cite{newman2010networks, jackson2010social} have emerged as one of the leading frameworks to describe complex social, economic and information systems.
In the last two decades, the network approach to complex systems has provided novel insights into various real-world problems, such as understanding the growth of information systems \cite{bianconi2001competition,kong2008experience,medo2011temporal}, identifying influential spreaders \cite{kitsak2010identification,ren2014iterative, lu2016vital}, and predicting the hitting time of an infectious disease \cite{brockmann2013hidden, iannelli2016effective}.
One of the central problems in network analysis is to assess whether an observed network property is a manifestation of a non-trivial phenomenon induced by the network's structure, a statistical consequence of the network's basic properties, or even a random fluctuation. For example, various community detection techniques \cite{fortunato2016community} -- such as the popular modularity optimization \cite{blondel2008fast,sobolevsky2014general} -- rely on quantifying how much the observed number of edges within a given set of nodes deviates from its expected value under a certain null model.
Network null models \cite{newman2001random,park2003origin,park2004statistical, squartini2011analytical, saracco2015randomizing, orsini2015quantifying, casiraghi2016generalized} serve this purpose by fixing one or more network properties while randomizing the rest.
These null models turn out to be essential for the detection of network organizational patterns such as communities \cite{blondel2008fast,mucha2010community}, rich clubs \cite{colizza2006detecting}, motifs \cite{milo2002network}, and nestedness \cite{jonhson2013factors}.

Despite the growing interest in the temporal evolution of complex networks \cite{medo2011temporal, holme2012temporal,rosvall2014memory, scholtes2014causality, holme2015modern, xu2016representing},
commonly used null models only focus on preserving structural network properties \cite{park2004statistical, orsini2015quantifying} and neglect the temporal patterns entirely.
For example, static network null models have been often applied to growing citation networks for the popular problems of detecting communities \cite{leicht2007large,takeda2010tracking, chen2010community,waltman2013smart} and classifying scientific papers into research fields \cite{waltman2012new}.

We show that omitting temporal information results in model-generated networks that exhibit highly unrealistic features, which in turn impairs their reliability as baselines to assess the significance of observed network properties.
To overcome this shortcoming, we introduce a time-aware methodology, which we jointly refer to as the \emph{dynamic configuration model} (DCM), which preserves not only the node degree sequence, but also the temporal linking patterns. This model allows us to randomize the system at an arbitrary temporal resolution, choose an appropriate resolution, and construct networks where each node's degree trajectory $k(t)$ is similar to that in the real network.
We use two datasets -- the citation network among the papers published by American Physical Society journals and the inspiration network of US movies (hereafter referred to as \emph{Papers} and \emph{Movies}, respectively) -- to show that differently from the static configuration model, the dynamic configuration model accurately reproduces temporal linking patterns of the real networks.

Differently from existing null models, by preserving the temporal linking patterns of the original network, 
the proposed model allows us to assess the significance of network structural properties also in 
settings where temporal patterns significantly affect structural measurements, which is the case for a wide range
of real systems  \cite{rosvall2014memory, scholtes2014causality, xu2016representing, chen2007finding,  mariani2015ranking, vidmer2015unbiased, vidmer2016essential}.
We apply the dynamic configuration model to three classes of network properties: (1) degree-degree correlations, (2) correlations between centrality metrics, (3) centrality metrics' ability to uncover significant nodes in the network.
We find that for the movie-movie citation network, the observed real properties can be largely explained by the dynamic configuration model.
This indicates that the movie citation network can be viewed as structurally random.
By contrast, the paper citation network is found to exhibit patterns that disappear when the network is randomized by the dynamic configuration model.
These patterns cannot be explained by existing network models, which calls for new mechanistic models of the citation network growth.
The provided results are examples; the proposed dynamic configuration model can be applied to assess the significance of any structural network property in growing networks.

\begin{figure*}
\centering
\includegraphics[scale=0.75,angle=0]{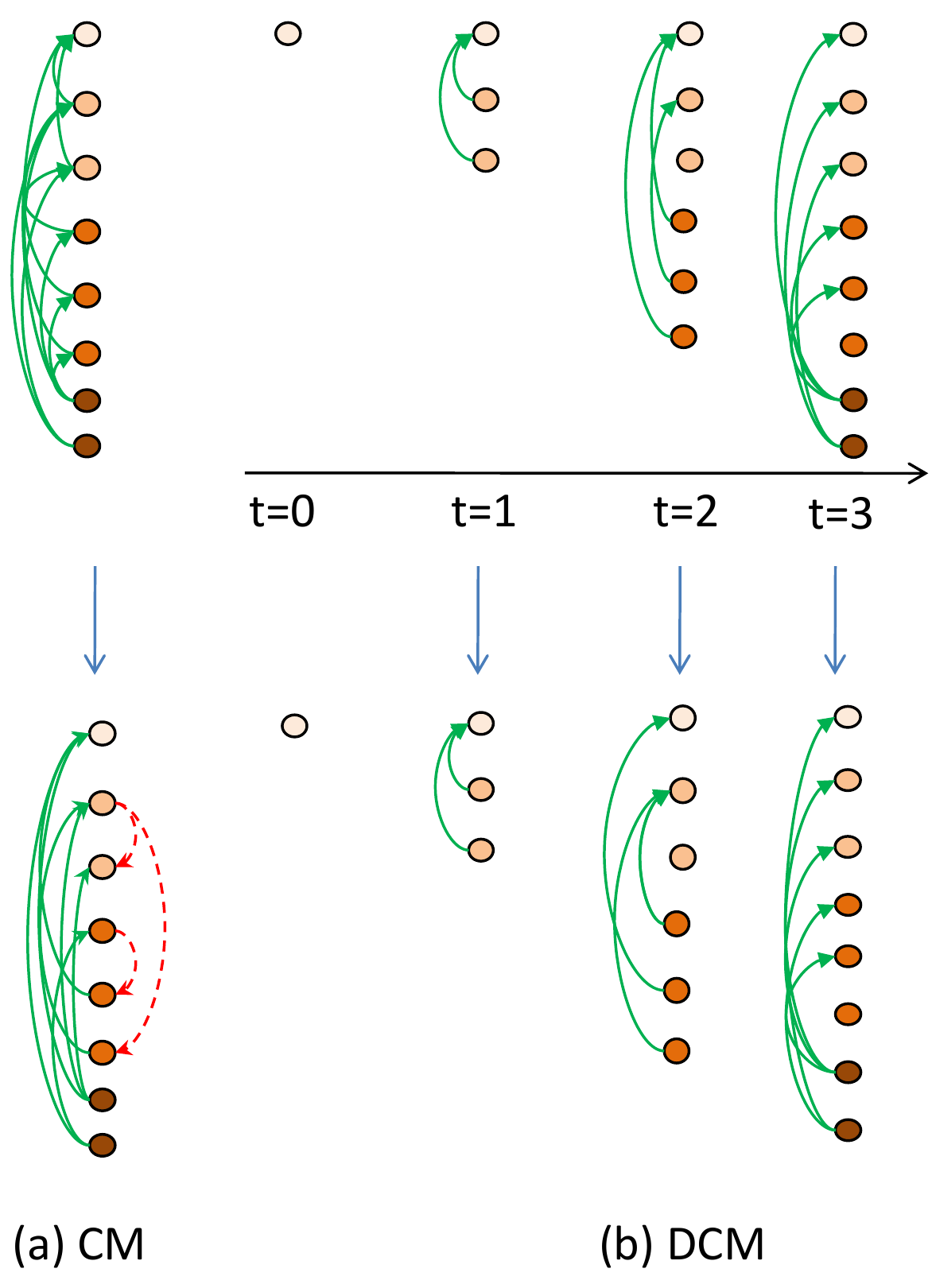}
\caption{(Color online) An illustration of how the Configuration Model (CM) and the Dynamic Configuration Model (DCM) operate in a growing network. Nodes are colored according to their age, from older (lighter) to more recent (darker). Edges that are directed forward in time are colored in red.}
\label{illustration}
\end{figure*}

\section{Network null models}
\label{sec:metrics}
A null model specifies the set of network properties to be kept fixed while randomizing the rest.
The classical configuration model (CM, paragraph \ref{sec:cm}) preserves the degree sequence $\{k^{in}_{i}, k^{out}_{i}\}$ where $k^{in}_{i}$ and $k^{out}_{i}$ are the in-degree and out-degree of node $i$.
Since the CM ignores the time structure of the network, the networks it produces exhibit unrealistic features (see paragraph \ref{sec:cm}), such as a substantial fraction of links pointing forward in time whereas in the two real networks studied here, all links point back in time.
By contrast, the dynamic configuration model (DCM, paragraph \ref{sec:dcm}) preserves not only the degree sequence, but also the nodes' degree trajectories. The networks produced with the DCM thus accurately reproduce the network's temporal linking patterns.

\subsection{Configuration model (CM)}
\label{sec:cm}
For a directed network, the (static) configuration model (CM) \cite{newman2001random,newman2010networks} generates directed random networks with given out- and in-degree sequences $\{k^{out}_i\}$ and $\{k^{in}_i\}$.
To this end, each node $i$ is endowed with $k^{out}_{i}$ outgoing-edge stubs and $k^{in}_{i}$ incoming-edge stubs.
A realization of the CM is formed by consecutively forming pairs of nodes with remaining stubs (always an out-stub with an in-stub) until there is no node with out- or in-going stubs left \cite{newman2001random}.
The random matching can generate self-loops and multiple edges 
However, for large networks they constitute only a small fraction of the total number of edges and therefore they can be safely discarded\footnote{In the two datasets analyzed here, we find that the fraction of discarded edges is of the order of $10^{-4}$.} \cite{newman2010networks}.
In this way, given a real directed network $\mathcal{G}$ and its degree sequences $\{k^{in}\}$ and $\{k^{out}\}$, one uses the CM to generate maximally randomized networks with the same in- and out-degree sequences as $\mathcal{G}$; the resulting randomized networks serve as a null model for patterns observed in the real network.
The randomized networks obtained with the CM from $\mathcal{G}$ are referred to as $\mathcal{G}$'s CM-randomized networks. 

By only preserving the individual nodes' degree values, the CM neglects the network's temporal patterns and, for this reason, can generate networks that exhibits highly unphysical temporal patterns. An illustration of this is provided in Fig. \ref{illustration}a: while only backward-directed edges (i.e., from more recent to older nodes) are allowed in a growing citation network, the CM-randomized networks can exhibit edges that point forward in time. This shortcoming of the CM is further discussed in Section \ref{sec:preserves} and motivates to introduce the Dynamic Configuration Model in the next section.

\subsection{Dynamic configuration model (DCM)}
\label{sec:dcm}
To amend the CM's ignorance of the time information, we introduce the dynamic configuration model (DCM).
We formulate the DCM for directed networks; adapting it to undirected networks is straightforward.
The DCM generates networks with fixed in- and out-degree time series (the final in- and out-degree values are thus automatically as the last points of the degree time series').
To this end, the system's time span $T$ is divided into $L$ temporal layers of equal duration $\Delta T=T/L$.  The in- and out-degree time series are defined as the sequences of in- and out-degree variations $\{\Delta k^{in}_{i,n}, \Delta k^{out}_{i,n}\}$, respectively, where $\Delta k^{out}_{i,n}$ and $\Delta k^{in}_{i,n}$ represent the change of in- and out-degree of node $i$, respectively, within the temporal layer $n$ ($n=1,\dots,L$).
A realization of the DCM is formed by individual temporal layers, where in layer $n$ we assign $\Delta k^{in}_{i,n}$ incoming and $\Delta k^{out}_{i,n}$ outgoing stubs to each node $i$, and match the in- and out-stubs at random.
Multiple edges and self-loops are discarded\footnote{The fraction of discarded edges is again of the order of $10^{-4}$.}.
Note that when $L=1$ (one layer limit), the DCM reduces to the CM. 

Given a real directed network and its degree time-series $\{\Delta k^{in}_{i,n}\}$ and $\{\Delta k^{out}_{i,n}\}$, one can use the DCM to generate a statistical ensemble of random networks with the same in- and out-degree time-series as the real network under consideration.
These networks then serve as a null model for both static and temporal patterns observed in the real network.
The randomized networks obtained with the DCM from $\mathcal{G}$ as referred to as $\mathcal{G}$'s DCM-randomized networks.
The number of temporal layers, $L$, is the sole parameter of the DCM model.

An illustration of how the DCM operates is provided in Fig. \ref{illustration}b. By dissecting the network's temporal evolution into layers and only performing within-layer randomizations of the network, the DCM allows us to obtain randomized networks that do not feature edges that violate the temporal constraint of the original network -- in the growing citation network shown in Fig. \ref{illustration}b, the randomized networks do not exhibit edges that travel forward in time. This aspect is further developed in the next Section. 

The expected number of edges $E(i\to j,n)$ from node $i$ to $j$ in the temporal layer $n$ is
\begin{equation}
E(i\to j,n)=\frac{\Delta k^{out}_{i,n}\, \Delta k^{in}_{j,n}}{\Delta E_n}
\label{expected_dcm}
\end{equation}
where $\Delta E_n$ is the number of edges introduced within the temporal layer $n$. Note that similarly to the CM \cite{park2003origin,squartini2011analytical}, Eq.~(\ref{expected_dcm}) cannot be used to estimate the probability $p(i\to j,n)$ that two nodes are connected as $E(i\to j, n)$ can be larger than one. While in this paper we compute the expected properties of the DCM-generated networks only numerically, extending the maximum-entropy framework by Squartini and Garlaschelli \cite{squartini2011analytical} to correctly estimate the probability that two nodes are connected with the DCM can lead to analytic computation of these properties.

The DCM is similar in spirit to the null model used in \cite{mucha2010community} for a complicated setting (a multilayer network with community structure). We focus here on the simplest possible setting of a growing directed network which is thus applicable to a broad range of systems. Generalizations to other settings (an undirected network, for example) are nevertheless possible.
The idea behind the DCM for growing networks is also reminiscent of the procedure used in \cite{squartini2011randomizing} to analyze the temporal evolution of the country-country International Trade Network (ITN), with important differences related to the different nature of the datasets. While Squartini et al. \cite{squartini2011randomizing} applied the CM to individual years of the ITN in order to study how network properties change with time, we applied the DCM to citation networks to assess whether the \emph{final} structural properties of the network are explained by the temporal evolution of each node's degree alone.
We emphasize that the partition of the nodes into
temporal layers is itself a key novel ingredient of the DCM which allows us to tune the temporal resolution of the analysis; how the chosen number of layers impact the properties of the randomized networks is discussed in the next subsection.

\begin{figure*}
\centering
\includegraphics[scale=0.8,angle=0]{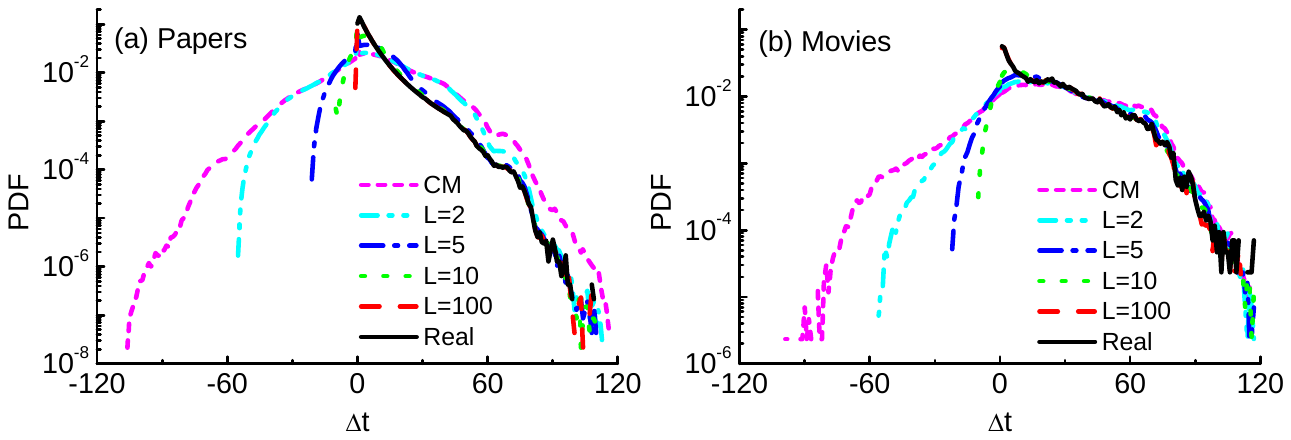}
\caption{(Color online) Distribution of the edge temporal lag $\Delta t$, where $\Delta t_{ij}=t_{i}-t_{j}$ for a directed edge $i\to j$, for \emph{(a) Papers} and \emph{(b) Movies}. Results are shown for the real data, the configuration model (CM), and the dynamic configuration model with various layer counts $L$.}
\label{Fig1a}
\end{figure*}

\begin{figure*}
\centering
\includegraphics[scale=0.8,angle=0]{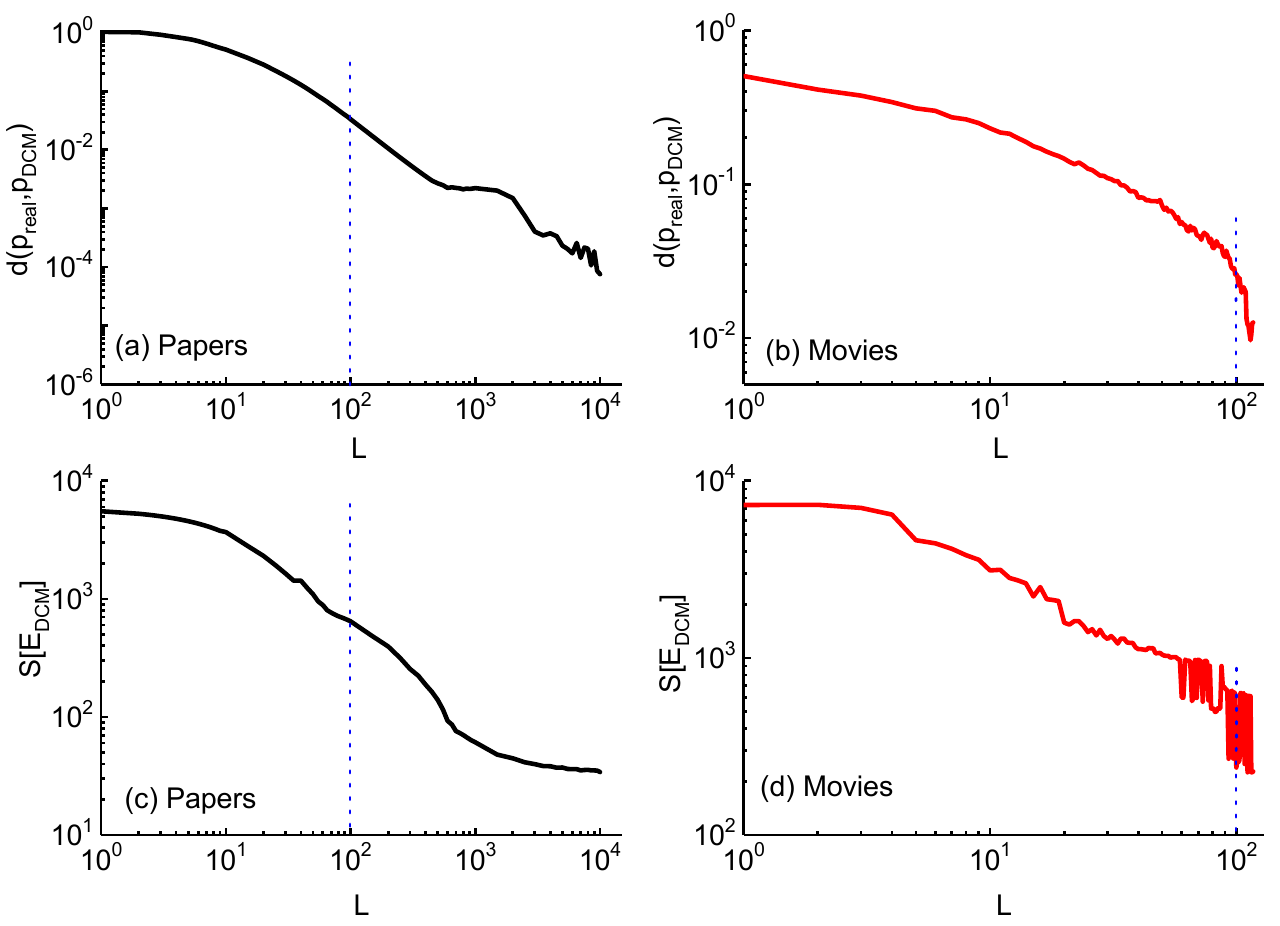}
\caption{(Color online) Distance $d(P_{real}, P_{DCM})$ between the time-lag distribution in the real and DCM-generated data for (a) \emph{Papers} and (b) \emph{Movies}. Entropy $S[E_{DCM}]$ of the dynamic configuration model as a function of the number of temporal layers $L$ ($L=1$ corresponds to the configuration model) for (c) \emph{Papers} and (d) \emph{Movies}. }
\label{Fig1b}
\end{figure*}

\begin{figure*}
\centering
\includegraphics[scale=0.8,angle=0]{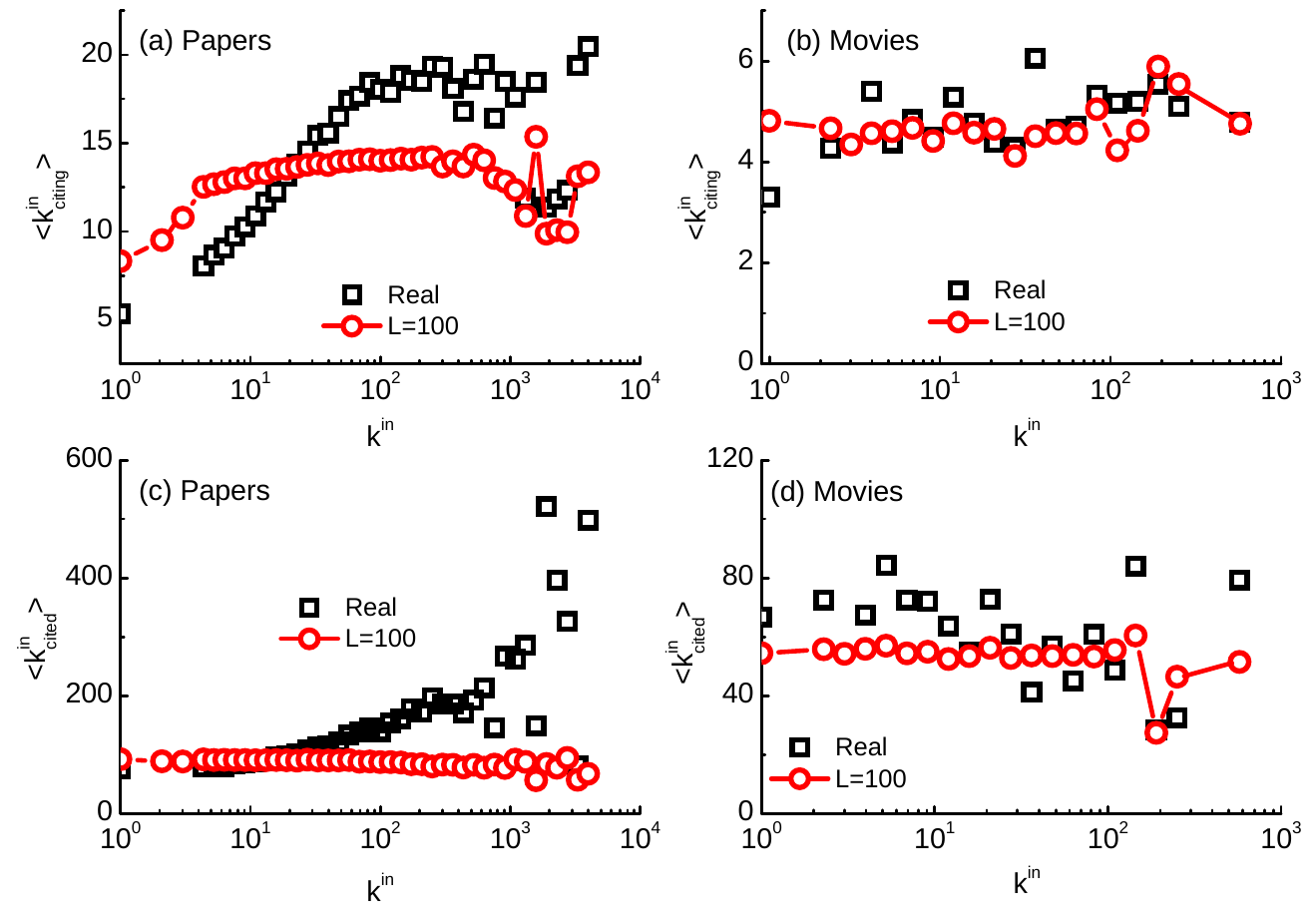}
\caption{(Color online) Relation between node indegree $k^{in}$ and the average indegree  $\braket{k^{in}_{citing}}$ of citing nodes for \emph{(a) Papers} and \emph{(b) Movies}.  Relation between node indegree $k^{in}$ and the average indegree  $\braket{k^{in}_{cited}}$ of cited nodes for \emph{(c) Papers} and \emph{(d) Movies}. Square and circle symbols represent the real data and the DCM-generated data, respectively.}
\label{Fig2}
\end{figure*}

\subsection{The DCM preserves real-data temporal linking patterns}
\label{sec:preserves}

Figure \ref{Fig1a}a,b compares the distribution of the edges' time lag in randomized networks obtained with the CM and the DCM with that found in the real data (see datasets' description in~\ref{datasets}). The \emph{time lag} of a directed edge $i\to j$ is defined simply as $t_{i}-t_{j}$, where $t_i$ denotes the time at which node $i$ enters the system (the paper publication time and the movie release time for \emph{Papers} and \emph{Movies}, respectively). In both real datasets, links always point back time; the time lag values are thus constrained to be positive. The CM networks show a much different pattern with a substantial fraction of links violating the original time ordering (In the CM, the fraction of forward links is  32\% and 18\% for \emph{Papers} and \emph{Movies}, respectively). This is a direct consequence of the CM's ignorance of the temporal dimension which is shared by a number of existing null models. Some links with negative time lag are produced also by the DCM but their fraction quickly diminishes as $L$ grows and the time lag distribution approaches to that of the real network.

To quantify the difference between the edge time lag distribution in real data, $P_{real}(\Delta t)$, and in DCM-randomized networks, $P_{DCM}(\Delta t)$, we calculate the $\mathbb{L}_1$-distance
\begin{equation}
d(P_{real}, P_{DCM})=\sum_t \big\vert P_{real}(t) - P_{DCM}(t)\big\vert
\end{equation}
where we uniformly divide the time-lag axis into bins of one-year duration and sum over all of them.
As shown in Figure~\ref{Fig1b}a,b, the distance $d(P_{real},P_{DCM})$ monotonously decreases with $L$. This is an expected result because the edge time lag error introduced by the DCM, which is at most $2\,\Delta T$, decreases with the number of layers. While this result might suggest that one should choose $L$ values as large as possible, large $L$ corresponds to short duration of individual layers which consequently leaves little space for randomness and limits the statistical significance of thus-obtained results. In the extreme case of temporal layers containing only one edge each, randomness has no place and the only possible DCM-randomized network is by definition identical with the input real network.
To quantify the level of randomness in the model-generated networks, we measure the entropy of the dynamic configuration model, defined as
\begin{equation}
S[E_{DCM}]=-\sum_{n=1}^{L}\sum_{i,j}\frac{E_{DCM}(i\to j,n)}{E_{n}}
\,\log\frac{E_{DCM}(i\to j,n)}{E_n}.
\end{equation}
which generally decreases with $L$ (Figs. 2c-d).

We set $L=100$ for both \emph{Papers} and \emph{Movies}, which avoids two extremes: unrealistic temporal patterns for too small $L$, and too small randomness for too large $L$. At the chosen value of $L$, $P_{DCM}$ matches well the real time-lag distribution and yields substantial model entropy.
It remains open whether one can devise a general statistically-grounded criterion to choose the value of $L$.
Nevertheless, the presented results obtained with the DCM do not alter qualitatively when $L$ changes, which suggests that the problem of finding an optimal value of $L$ is not essential for practical purposes. Our results are based on ten independent realizations of the DCM.

An alternative time-respecting model based on layers composed of an equal number of nodes instead of layers of equal temporal duration is studied and compared with the DCM in \ref{ndcm}.

\section{Using the DCM to assess the significance of observed network properties}
In this section, we apply the DCM to assess the significance of three distinct network properties: (1) degree-degree correlations, (2) correlations between node centrality metrics, and (3) performance of node centrality metrics in identifying significant nodes.

\begin{figure*}
\centering
\includegraphics[scale=0.8,angle=0]{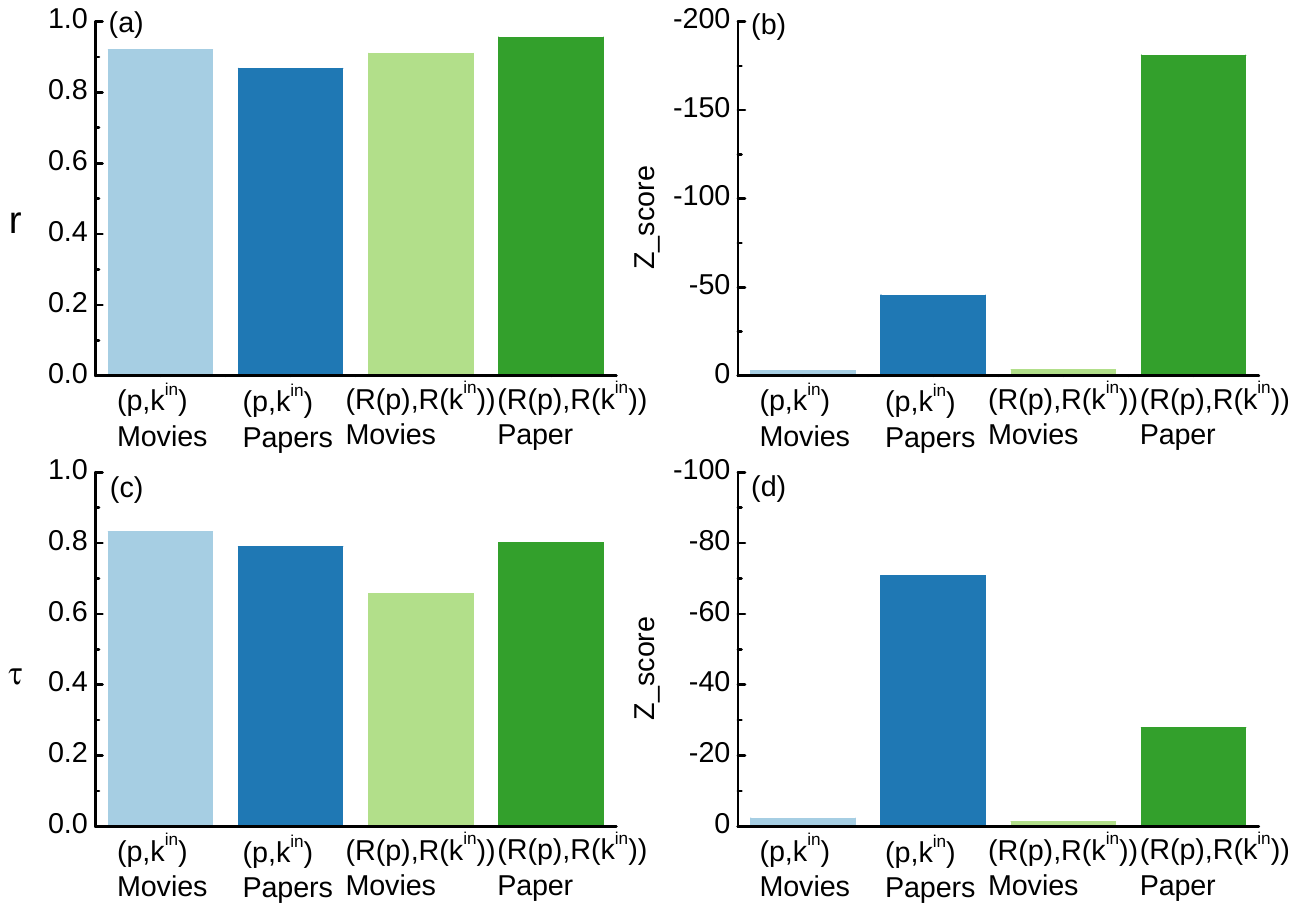}
\caption{(Color online) \emph{(a)} Pearson's correlations $r$ between the studied metrics for \emph{Papers} and \emph{Movies}, and \emph{(b)} associated $z$-scores obtained with the DCM.
\emph{(c)} Kendall $\tau$ values between the studied metrics for \emph{Papers} and \emph{Movies}, and \emph{(d)} associated $z$-scores obtained with the DCM.}
\label{Fig3}
\end{figure*}

\subsection{Degree-degree correlations}
\label{deg_deg_corr}
The degree-degree correlation is usually visualized by the assortativity plot \cite{pastor2001dynamical}
which displays the average degree of a node's neighbors as a function of the node degree.
Out of the various possible options for degree-degree correlation in directed networks~\cite{foster2010edge}, we focus here on two distinct cases: the indegree-indegree dependence between a node and the nodes it points to (cited nodes), as well as the nodes it is pointed by (citing nodes).

The \emph{Papers} network exhibits a clear assortative pattern in both cases and this pattern cannot be explained by the DCM (see Fig.~\ref{Fig2}a, c). Note that the possible interpretations of these two cases are different: while panel (a) suggests that little cited papers are cited by other little cited papers,
panel (b) suggests that the authors of highly cited papers choose and cite other highly cited papers.
By contrast, no significant indegree-indegree correlation are found in the \emph{Movies} network and the same is true for its DCM-randomized networks.

Degree-degree correlations impact, among others, the correlations between different node centrality metrics. For uncorrelated networks, for example, PageRank score is on average expected to be proportional to indegree (in other words, PageRank score carries no more information than node indegree). A similar result holds for the node $H$-index introduced in \cite{lu2016h}: the $H$-index of a node is highly correlated with node indegree for uncorrelated networks \cite{pastor2016topological}. We shall discuss the implications of the DCM on the indegree-PageRank relation in the following sections.

\subsection{Relations between node centrality metrics}
\label{comparison}
While prior studies \cite{mariani2015ranking, upstill2003predicting, chen2004local,pandurangan2006using, fortunato2006approximating} have reported the values of indegree-PageRank correlations measured in real data, whether the observed correlations are large or small is often discussed without any reference to a suitable null model. We use here DCM-randomized networks to assess whether correlations between network metrics of structural importance can be explained by degree dynamics or not.

Consider the vectors $\mathbf{s}_1$ and $\mathbf{s}_2$ of scores produced by two different metrics, and suppose that we are interested in the linear Pearson correlation $r(\mathbf{s}_1,\mathbf{s}_2)$ between the two score vectors. We estimate the significance of $r(\mathbf{s}_1,\mathbf{s}_2)$ in a given network $\mathcal{G}$ by computing its $z$-score with respect to its distribution in $\mathcal{G}$'s DCM-randomized networks as follows
\begin{equation}
z[r(\mathbf{s}_1,\mathbf{s}_2)]=\frac{r(\mathbf{s}_1,\mathbf{s}_2)-\mu[r(\mathbf{s}_1,\mathbf{s}_2)]}{\sigma[r(\mathbf{s}_1,\mathbf{s}_2)]}.
\end{equation}
Here $\mu[r(\mathbf{s}_1,\mathbf{s}_2)]$ and $\sigma[r(\mathbf{s}_1,\mathbf{s}_2)]$ represent the mean and the standard deviation, respectively, of $r(\mathbf{s}_1,\mathbf{s}_2)$ over the ensemble of DCM-randomized networks. An analogous definition is used to estimate the significance of Kendall's tau correlation $\tau(\mathbf{s}_1,\mathbf{s}_2)$.

In addition to node indegree $k^{in}$ and PageRank score $p$, we study here also age-rescaled variants of these two metrics, rescaled indegree $R(k^{in})$ and rescaled PageRank $R(p)$, which were proposed in~\cite{mariani2016identification} (see \ref{centrality} for the definition of the four metrics and computation details). The main idea behind the rescaled metrics is that the rescaling to a large extent removes the strong age bias of the original metrics and thus makes it possible to compare the nodes regardless of their age. The procedure described here can be applied to assess the significance of the relation between any other pair of node centrality metrics.

Pearson and Kendall correlation between the centrality metrics are high in both \emph{Movies} and \emph{Papers} network (see Figure~\ref{Fig3}a,c). The $z$ score values in Figure~\ref{Fig3}b,d imply that the interpretation of these similar correlation values is actually different between the two studied systems. In the \emph{Movies} network, the correlation observed in real datasets matches well (the $z$ score is close to zero) the correlation observed in DCM-randomized networks. This indicates that the information conveyed by indegree is the same as that conveyed by PageRank and the existing discrepancies (manifested by the correlation lower than one) can be explained by random fluctuations. In the \emph{Papers} network, the correlation observed in real datasets is \emph{significantly lower} (the $z$ score is strongly negative) than in DCM-randomized networks, which indicates that PageRank scores carry additional information that is not captured by node indegree. We illustrate an interesting consequence of this result in the following section. In summary, we find that similar correlation values can yield vastly different $z$ scores when compared with DCM-randomized networks. This indicates that null models are essential for a proper interpretation of measurements in complex networks.

\begin{figure*}
\centering
\includegraphics[scale=0.8,angle=0]{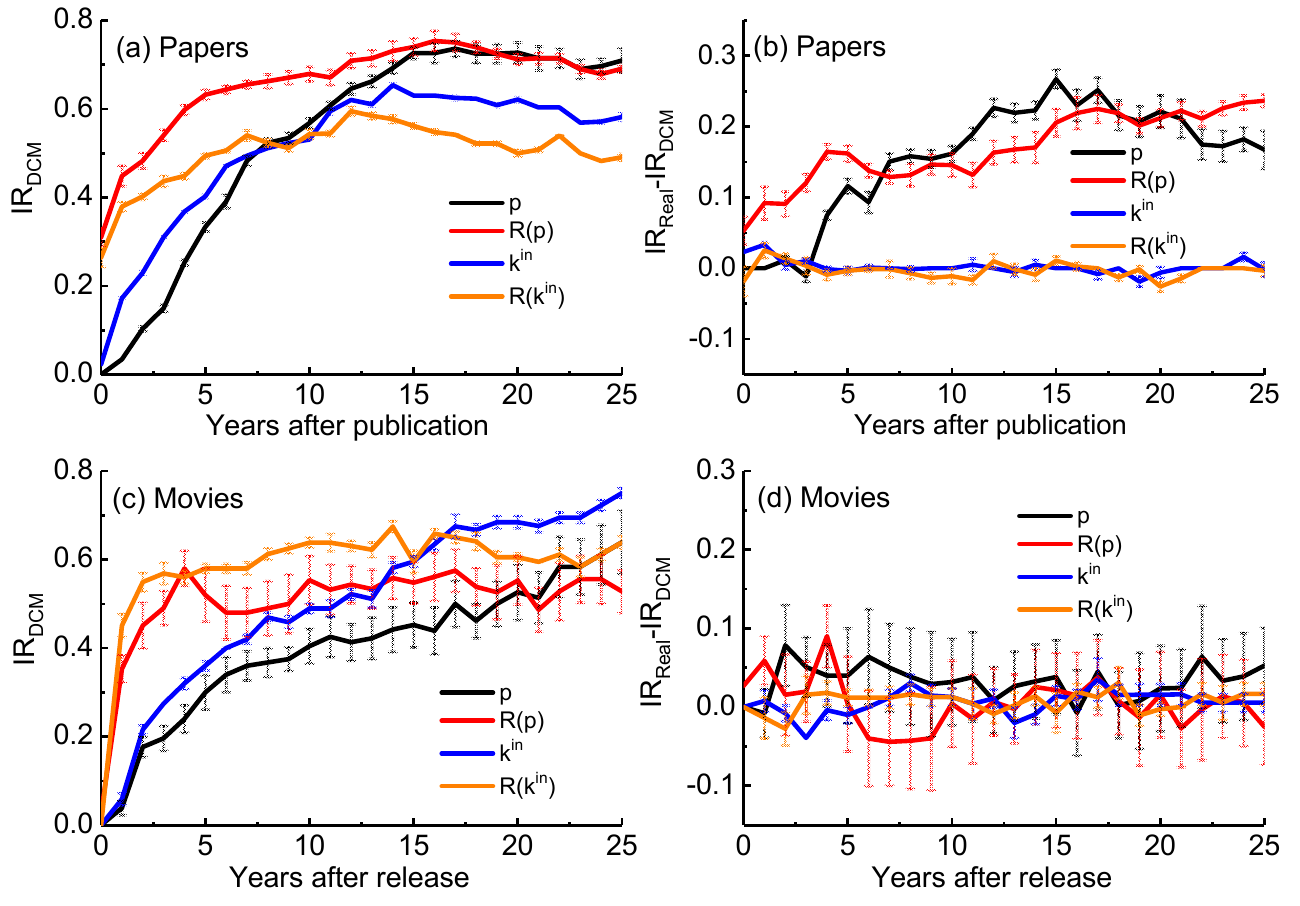}
\caption{(Color online) Identification rate as a function of paper age \emph{(a)} for the Milestone Letters in \emph{Papers} and \emph{(c)} for the Oscar-awarded movies in \emph{Movies}.
Identification-rate difference between the original dataset and the DCM-generated networks for \emph{(b) Papers} and \emph{(d) Movies}. The curves represent averages over ten independent realizations of the randomization process; the error bars represent the standard errors. 
}
\label{Fig4}
\end{figure*}

\subsection{The performance of centrality metrics in identifying significant nodes}
\label{identification}
This section discusses the implications of the previous findings on the ability of indegree and PageRank to identify significant nodes. In a recent work~\cite{mariani2016identification}, some of the authors of this manuscript used the \emph{Papers} network to evaluate the ranking of nodes produced by various node centrality metrics with a particular emphasis on the ranking positions of a set of fundamental papers, called Milestone Letters. These were chosen by the Physical Review Letters editors for their ``long-lived contributions to physics, either by announcing significant discoveries, or by initiating new areas of research''. Differently from the common static evaluation of bibliometric indicators \cite{chen2007finding, sidiropoulos2006generalized, dunaiski2016evaluating, medo2016model}, the analysis presented in~\cite{mariani2016identification} takes the temporal dimension into account and concerns the ability of different metrics to single out the milestone papers as a function of their age (the logic behind this is that a good ranking method should be able to rank a milestone paper high short after it has been published).

Here, we use the DCM to deepen that result and show that the observed performance gap between network-based indicators (PageRank $p$ and time-rescaled PageRank $R(p)$) and indicators based on citation count (citation count $k^{in}$ and rescaled citation count $R(k^{in})$) disappears when we randomize the network and thereby destroy the real network's topological patterns.

\subsubsection{Identification of milestone papers in the Papers network}
The ranking performance of a metric is measured by the fraction of milestone papers that appear in top 1\% of the ranking $t$ years after their publication (see~\cite{mariani2016identification} for details); this quantity is referred to as the \emph{identification rate}.
The identification rate achieved by the four considered metrics in the \emph{Papers} network is shown in Figure~\ref{Fig4}a (this panel is identical with the result shown in~\cite{mariani2016identification}).
The network-based indicators ($p$ and $R(p)$) significantly outperform local metrics ($k^{in}$ and $R(k^{in})$) in identifying the milestone papers. Thanks to the suppression of PageRank's time bias that generally favors old papers, rescaled PageRank is superior to PageRank until approximately 15 years after publication; from then on, the two metrics perform similarly.

We now use the DCM to assess whether the same holds in DCM-randomized networks.
To this end, Figure~\ref{Fig4}b shows the identification rate difference between the real and DCM-randomized networks. We see that the observed differences are small for both $R(k^{in})$ and $k^{in}$. This is a directed consequence of the fact that the DCM preserves the citation time-series of all the papers. By contrast, $p$ and $R(p)$ perform substantially better in the real network than in the DCM-randomized networks. This implies that the dynamics of the paper citation count alone cannot
explain the superior performance of $R(p)$ and $p$ in identifying the milestone papers \cite{mariani2016identification}.

\subsubsection{Identification of awarded movies in the \emph{Movies} network}
Data on expert-based assessment of movie significance are available as well and have been used, for instance, to assess whether classification algorithms based on centrality metrics can identify expert-selected movies of lasting importance \cite{wasserman2015cross}.
We carry out the same identification rate analysis as before with the movies that received at least five Oscar awards playing the same ``ground truth'' role as we previously assigned to the Milestone Letters.
Figure~\ref{Fig4}c shows that the rescaled metrics are again superior to the unrescaled ones short after the movie release.
However, the performance differences shown in Figure~\ref{Fig4}c are close to zero which indicates that the four metrics behave in the real network in almost the same way as they do in the DCM-randomized networks.
This suggests that in the \emph{Movies} network, there is no gain in using the whole network to compute node centrality.

To summarize, whether PageRank-related metrics outperform indegree-related metrics depends on topological properties of the network under consideration.
In particular, if the given network is uncorrelated, PageRank-related and indegree-related metrics yield similar information and there is thus no gain in assessing node importance using PageRank which is furthermore computationally more demanding than indegree.
If instead the network exhibits non-trivial degree-degree correlations, indegree-PageRank correlation is smaller than the expected value in an uncorrelated network, and PageRank-related methods have the potential
to produce node rankings that are superior to those obtained with indegree-related metrics.

\section{Discussion}
\label{sec:conclusions}

We have introduced here a network null model, called dynamic configuration model, which preserves not only the network's degree sequence, but also the degree time series of individual nodes. In the same way as the configuration model generates random networks with arbitrary degree sequence \cite{newman2001random}, the dynamic configuration model produces random networks with arbitrary degree dynamics. The failure of the configuration model in matching the real networks' temporal patterns undermines the effectiveness of its use as a model to estimate the expected values of structural quantities in growing networks. This has deep implications \cite{zeng2017community} for the problem of community detection \cite{fortunato2016community}, for example, where the classical modularity--optimization algorithm \cite{blondel2008fast} uses the configuration model to estimate the expected number of edges between pairs of nodes\footnote{The modularity--optimization algorithm has been applied to the citation network of scientific papers in \cite{leicht2007large,chen2010community,takeda2010tracking,vsubelj2016clustering}, among others.}.
Differently from the static configuration model, the new model is able to accurately reproduce the original network's temporal patterns, thus providing an improved baseline to assess the significance of observed properties of growing networks.
While we have focused on growing citation networks here, an open challenge for future research is to determine how best to randomize different kinds of networks that exhibit temporal patterns (e.g. networks where nodes can create outgoing edges at multiple times \cite{perra2012activity,mariani2015ranking}, and temporal networks \cite{holme2012temporal,holme2015modern}).

We stress that in this work we used the dynamic configuration model to numerically estimate the expected properties of model networks based on the randomization procedure introduced in \cite{newman2001random}. By contrast, a stream of literature \cite{park2004statistical,squartini2011analytical,fagiolo2013null} has focused on analytically computing the expected properties of model networks. In particular, the maximum-entropy approach by Squartini and Garlaschelli \cite{squartini2011analytical} allows one to correctly estimate the connection probability between pairs of nodes in the configuration model, which can be used in turn to analytically compute expected network structural properties. Extending analytical methods to the dynamic configuration model is an important open direction for future research.

Among others, three additional extensions of our work are possible. First, in a similar spirit as the $dk$-series introduced in \cite{orsini2015quantifying} generalizes the original configuration model, one could try to construct time-respecting null models that preserve not only the individual nodes' time series, but also the dynamics of higher-order structural properties, such as degree-degree correlations.
Second, the proposed null model can be extended to include deletion of edges in a similar spirit as the random graph for temporal networks recently proposed in ref. \cite{zhang2016random}.
Third, a null model that reflects the network evolution better than existing null models can be used to improve community detection in evolving networks.
This will be addressed in a forthcoming work \cite{zeng2017community}.

Our results on the citation network of scientific papers can be also viewed from another perspective. From its definition, it follows that the DCM applied on citation data can be interpreted as a model where scientists mindlessly follow the current trends when choosing which paper to cite.
The fact that the real data exhibit structural properties substantially different with respect to those found in the DCM-randomized networks can be interpreted as an encouraging sign that the citing behavior of scientists is to a considerable extent different from simply picking the currently trending papers.
This directly challenges the existing growth models for academic citation networks that assume simpler mechanisms of how to choose which papers to cite \cite{medo2011temporal, wang2013quantifying}.
Our findings
call for more complex models where more complicated and possibly also diversified citation strategies exist in the system.
It remains open whether to the observed non-trivial patterns can be reproduced by a unipartite network model -- possibly an extension of the existing models \cite{medo2011temporal, wang2013quantifying} -- or if additional layers of complexity (such as the behavior of authors, research groups, and institutions) will need to be introduced in the modeling framework.

\appendix

\section{The datasets}
\label{datasets}
\paragraph{Papers.}
We denote as \emph{Papers} the citation network composed of $E=4,672,812$
directed edges between the
$N=449,935$ papers published in American Physical Society journals between $1893$ and $2009$.
The data can be obtained under request at \url{http://journals.aps.org/datasets}.
Edges are time-stamped with the temporal resolution of one day.

\paragraph{Movies.}
We denote as \emph{Movies} the citation network composed of $E=42,794$ directed edges between
$N=15,425$ that compose the giant component of the network of US movies released between $1894$ and $2011$.
The data is publicly available at \url{https://amaral.northwestern.edu/resources/data-sets/us-film-citation-network},
were extracted by the authors of ref. \cite{wasserman2015cross} from the movie reference data available at
\url{ftp://ftp.fu-berlin.de/pub/misc/movies/database/}.
Edges are time-stamped with the temporal resolution of one year.




\begin{figure*}
\centering
\includegraphics[scale=0.8,angle=0]{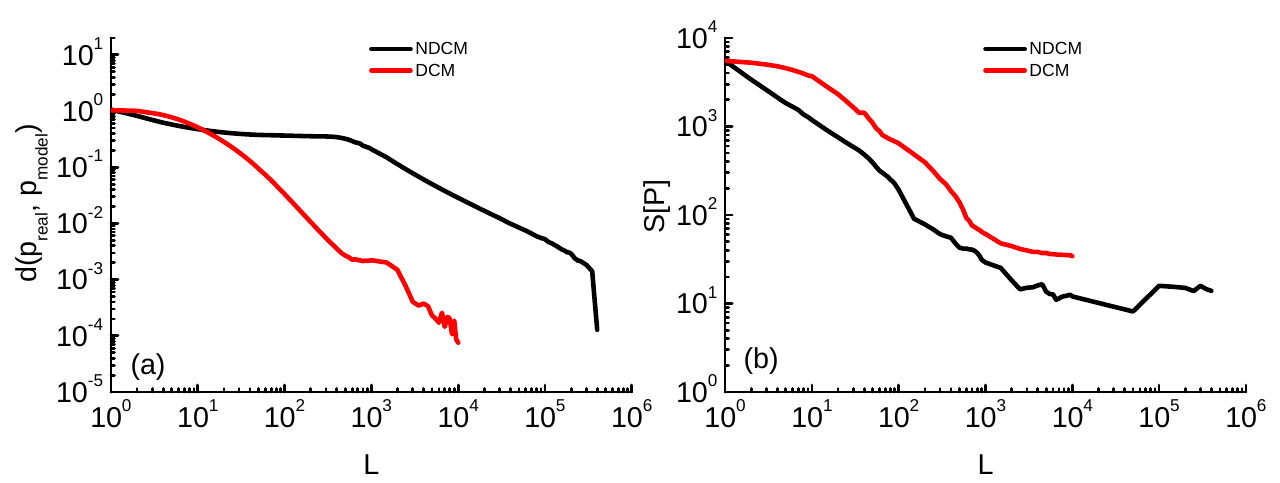}
\caption{(Color online) (a) Distance $d(P_{real}, P_{model})$ between the time-lag distribution in the real and model-generated data for \emph{Papers}, for both the DCM and the NDCM. (b) Entropy $S[E_{model}]$ of the model as a function of the number of temporal layers $L$ ($L=1$ corresponds to the configuration model) for \emph{Papers}, for both the DCM and the NDCM. }
\label{Fig-NDCM}
\end{figure*}

\begin{figure*}
\centering
\includegraphics[scale=0.8,angle=0]{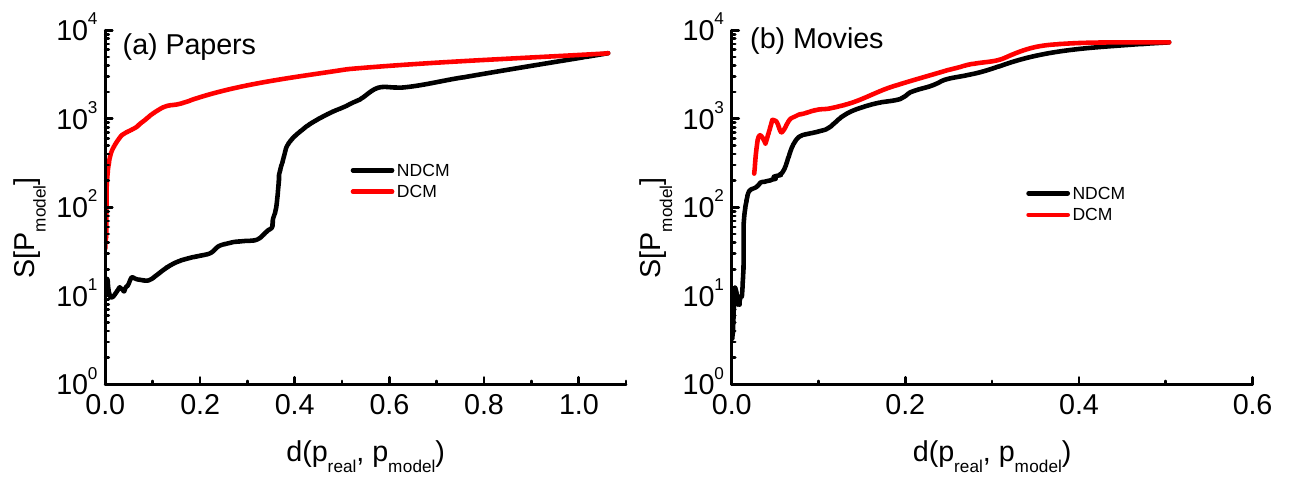}
\caption{(Color online) Relation between distance and entropy for the DCM and the NDCM, for (a) \emph{Papers}; (b) \emph{Movies}.}
\label{Fig-NDCM2}
\end{figure*}

\section{Node-based dynamic configuration model (NDCM)}
\label{ndcm}
We tested an alternative time-respecting null model, which we refer here as node-based dynamic configuration model (NDCM), based on the number of nodes instead of real time.
This means that we sorted the $N$ nodes according to their age, and we divided them into $L$ layers composed of the same number $\Delta N=N/L$ of nodes. 
The temporal duration of each layer is therefore not constant as in the DCM, but it is given by the difference between the publication times of the most recent and the oldest nodes that belong to the layer.
Similarly to the DCM, within each layer $n$, each node $i$ is endowed with $\Delta k_{i,n}^{out}$ and $\Delta k^{in}_{i,n}$ out-going and in-coming stubs, respectively, where $\Delta k_{i,n}^{out}$ and $\Delta k^{in}_{i,n}$ represent node variation of outdegree and indegree, respectively, within the time frame that corresponds to layer $n$.

The results for the NDCM are qualitatively in agreement with those obtained for the DCM. In particular
\begin{itemize}
\item The accuracy of the model in matching the real networks' temporal patterns improves with the number of layers (see Fig. \ref{Fig-NDCM}, left). 
\item The entropy of the model decreases with the number of layers (see Fig. \ref{Fig-NDCM}, right).
\end{itemize}
However, with respect to the DCM, the NDCM requires a larger number of layers to achieve the same accuracy as the DCM (Fig. \ref{Fig-NDCM2}). For example, to achieve the same accuracy achieved by the DCM with $L_{DCM}=100$ ($d(P_{DCM},P_{real})=0.0335$ for \emph{Papers}), we need a NDCM model with many more layers ($L_{NDCM}=8000$) and much smaller entropy ($S(P_{NDCM})=12.09$ as opposed to $S(P_{DCM})=646.03$). 

The larger entropy of the DCM implies that with respect to the NDCM, the DCM is able to explore a larger ensemble of networks compatible with the given temporal constraints. For this reason, we prefer to randomize the networks with he DCM in the main text.

\section{Results as a function of the number of layers}
\label{differentL}
In the main text, we have studied how changing the number $L$
of layers affects the properties of the DCM-generated networks.
In particular, increasing $L$ improves the DCM's accuracy in reproducing real networks' temporal patterns but, at the same time, increases the number of constraints on the randomized networks and, as a result, leads to a smaller ensemble of random networks and to a lower model entropy.

In the main text, we set $L=100$ which produces random networks that exhibit temporal patterns in good agreement with the real patterns (Fig. \ref{Fig1a}) and, at the same time, has entropy significantly larger than zero (Fig. \ref{Fig1b}). In this Appendix, we show that the conclusions drawn with the DCM with $L=100$ are robust with respect to other choices of $L$.

\begin{figure*}
\centering
\includegraphics[scale=0.8,angle=0]{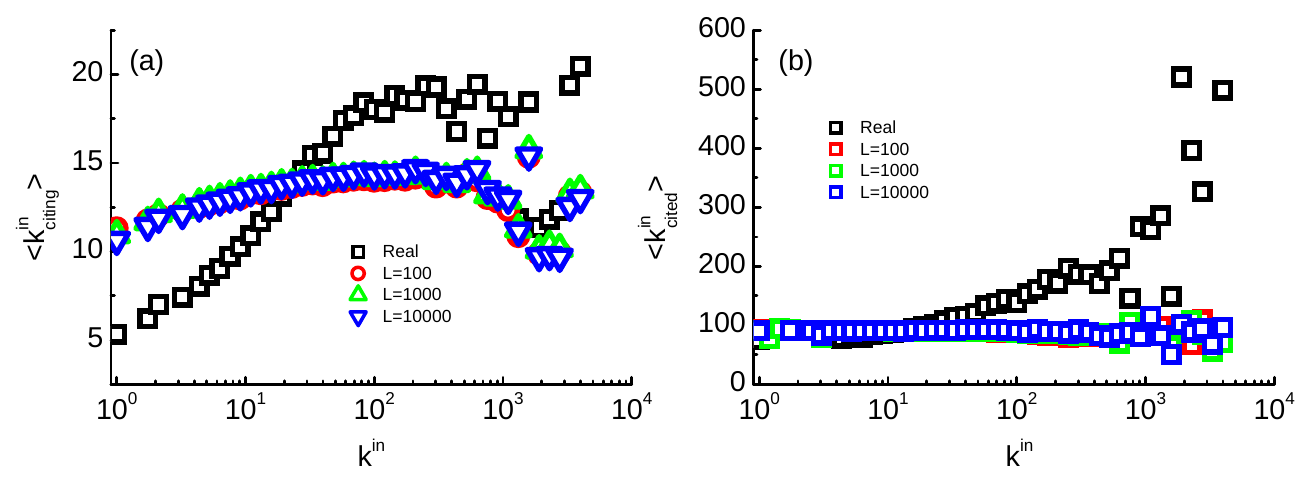}
\caption{(Color online) Relation between node indegree $k^{in}$ and the average indegree  $\braket{k^{in}_{citing}}$ of citing nodes (left) and relation between node indegree $k^{in}$ and the average indegree  $\braket{k^{in}_{cited}}$ of cited nodes (right) for \emph{Papers}. Square and circle symbols represent the real data and the DCM-generated data (with different values of $L$), respectively.}
\label{assortativity_appendix}
\end{figure*}

\begin{figure*}
\centering
\includegraphics[scale=0.8,angle=0]{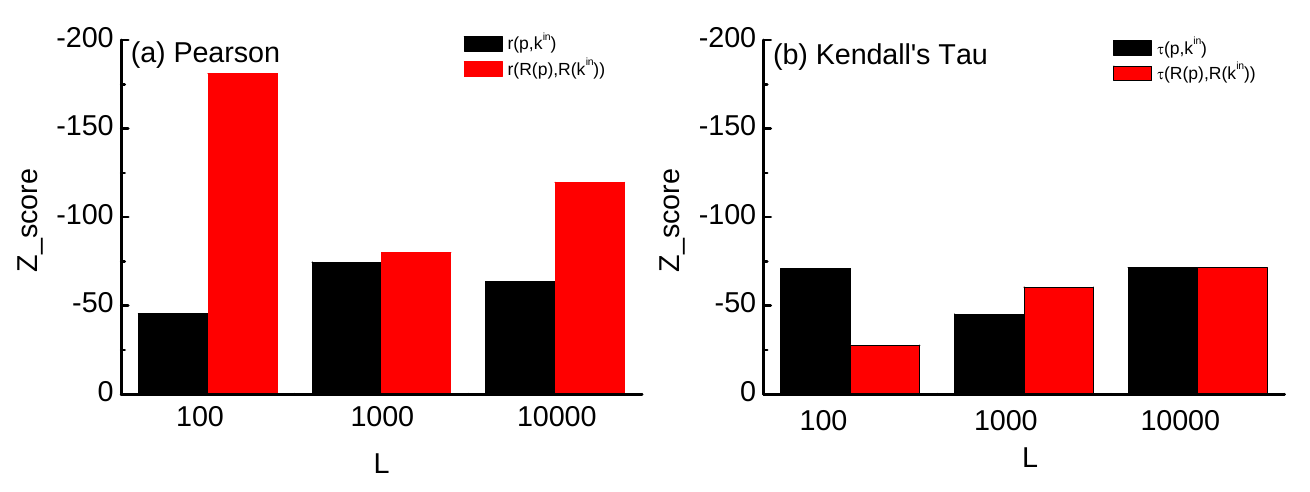}
\caption{(Color online) Results for \emph{Papers} obtained with the DCM with different values of $L$. \emph{(a)} $z$-scores for the Pearson's correlations $r$ between the various metrics, and \emph{(b)} $z$-scores for the Kendall $\tau$ values between the various metrics.}
\label{correlations_appendix}
\end{figure*}

Fig. \ref{assortativity_appendix} shows the assortativity plots for \emph{Papers} for different values of $L$. We only consider values of $L$ larger than $100$, i.e., values for which the randomized networks' temporal patterns match the real pattern better than with $L=100$. The Figure shows that the assortativity plots are little sensitive to the choice of $L$.
Fig. \ref{correlations_appendix} shows the $z$-score values for the metrics' correlations for different $L$ values. With respect to the values chosen in the main text, the $z$-score values are different yet much larger than one in modulus. This confirms that the significance of the observed correlations are also detected by the DCM with different choices of $L$ ($L=1000, 10000$).

\section{Network centrality metrics}
\label{centrality}
\paragraph{Indegree} The indegree of a given node is defined as the number of incoming edges received by that node. In terms of the network's adjacency matrix $\mathbf{A}$ (in a directed network, $A_{ij}=1$ if node $j$ points to node $i$, $A_{ij}=0$ otherwise), the indegree $k^{in}_i$ of a node can be simply expressed as $k_{i}^{in}=\sum_j A_{ij}$.

\paragraph{PageRank} The PageRank vector of scores $\mathbf{p}$ is defined by the following equation
\begin{equation}
 \mathbf{p}=\alpha \,\mathbf{P}\,\mathbf{p}+(1-\alpha)\mathbf{v},
\label{pr_eq}
 \end{equation}
where $P_{ij}=A_{ij}/k^{out}_j$ and $\mathbf{v}$ is a uniform teleportation vector ($v_{i}=1/N$ for all nodes $i$).
We set here $\alpha=0.5$ which is the usual choice in citation networks \cite{walker2007ranking}.
Eq.~(\ref{pr_eq})can be interpreted as the stationary equation of a stochastic process on the network where a random walker
either follows the network's edges
with probability $\alpha$, or he jumps to a randomly chosen node with probability $1-\alpha$.

\paragraph{Rescaled indegree and rescaled PageRank}
Rescaled indegree $R(k^{in})$ and rescaled PageRank $R(p)$ aim to suppress the temporal bias of indegree and PageRank, respectively
\cite{newman2014prediction, mariani2016identification}.
For both metrics, each node's structural centrality score is only compared with the scores of nodes of similar age.
Node $i$'s rescaled indegree score $R_i(k^{in})$ quantifies the number of standard deviations node $i$ outperforms with respect to
nodes of similar age with respect to $k^{in}$.
In formulas,
\begin{equation}
 R_i(k^{in})=\frac{k_{i}^{in}-\mu_{i}(k^{in})}{\sigma_{i}(k^{in})},
\end{equation}
where $\mu_{i}(k^{in})$ and $\sigma_{i}(k^{in})$ represent mean value and standard deviation, respectively,
of the indegree of the nodes $j\in[i-\Delta_k/2,i+\Delta_k/2]$.
Analogously, the rescaled PageRank score $R_i(p)$ of node $i$ is defined as \cite{mariani2016identification}
\begin{equation}
 R_i(p)=\frac{p_i-\mu_{i}(p)}{\sigma_{i}(p)},
\end{equation}
where $\mu_{i}(p)$ and $\sigma_{i}(p)$ represent mean value and standard deviation, respectively,
of the indegree of the nodes $j\in[i-\Delta_p/2,i+\Delta_p/2]$.

For \emph{Papers}, we set $\Delta_p=\Delta_c=1000$ \cite{mariani2016identification}.
For \emph{Movies}, we set $\Delta_p=\Delta_c=500$. A potential issue for \emph{Movies} is that
we do not know the exact order of movies released in the same year due to the time resolution of the data.
For this reason, we randomize the order of movies published in the same year in order to assign the temporal window. We observed that our findings do not depend on the particular order of movies released in the same year; for simplicity, the results presented in this article refer to one particular realization of the order of movies released in the same year.

\section*{Acknowledgements}
We wish to thank Giacomo Vaccario for inspiring discussions and useful suggestions. This work was supported by the EU FET-Open Grant No. 611272 (project Growthcom).

The authors declare that they have no competing financial interests. Correspondence and requests for materials should be addressed to M. S. Mariani~(email: manuel.mariani@unifr.ch).

\providecommand{\newblock}{}

\end{document}